\documentstyle[12pt,epsfig,axodraw]{article}

\newcommand{\bea}{\begin{eqnarray}}
\newcommand{\eea}{\end{eqnarray}}

\newcommand{\bcen}{\begin{center}}
\newcommand{\ecen}{\end{center}}

\newcommand{\lsim}{\raisebox{-0.13cm}{~\shortstack{$<$ \\[-0.07cm] $\sim$}}~}
\newcommand{\gsim}{\raisebox{-0.13cm}{~\shortstack{$>$ \\[-0.07cm] $\sim$}}~}

\newcommand{\ra}{\rightarrow}
\newcommand{\ee}{e^+e^-}
\newcommand{\s}{\\ \vspace*{-3mm} }
\newcommand{\nn}{\noindent}
\newcommand{\non}{\nonumber}
\newcommand{\beq}{\begin{eqnarray}}
\newcommand{\eeq}{\end{eqnarray}}

\newcommand{\tb}{\tan\beta}

\newcommand{\ct}[1]{c_{\theta_#1}}
\newcommand{\st}[1]{s_{\theta_#1}}
\newcommand{\sinb}{\sin\beta}
\newcommand{\cosb}{\cos\beta}

\def\t1{\tilde{t_1}}

\begin{document}

\def\thefootnote{\fnsymbol{footnote}}

\begin{flushright}
PM/99--29\\
\end{flushright}

\vspace{1cm}

\begin{center}

{\large\sc {\bf Decays of the Lightest Top Squark}}

\vspace{1cm}

{\sc C. Boehm, A. Djouadi and Y. Mambrini} 

\vspace{0.5cm}

Laboratoire de Physique Math\'ematique et Th\'eorique, UMR5825--CNRS,\\
Universit\'e de Montpellier II, F--34095 Montpellier Cedex 5, France.

\end{center}

\vspace{2cm}

\begin{abstract}

\nn 
We analyze higher order decay modes of the lightest top squark $\tilde{t}_1$ 
in the Minimal Supersymmetric extension of the Standard Model (MSSM), where 
the lightest SUSY particle (LSP) is assumed to be the neutralino $\chi_1^0$. 
For small $\tilde{t}_1$ masses accessible at LEP2 and the Tevatron, we show 
that the four--body decay mode into the LSP, a bottom quark and two massless 
fermions, $\tilde{t}_1 \ra b \chi_1^0 f\bar{f}'$, can dominate in a wide range 
of the MSSM parameter space over the loop--induced decay into a charm quark 
and the LSP, $\tilde{t}_1 \ra c \chi_1^0$. This result might affect the 
experimental searches on this particle, since only the later signal has
been considered so far. 
\end{abstract}

\newpage

\def\thefootnote{\arabic{footnote}}
\setcounter{footnote}{0}

\subsection*{1. Introduction}

Supersymmetric (SUSY) theories, and in particular the Minimal Supersymmetric 
extension of the Standard Model (MSSM) \cite{R1,R2}, predict the existence of a 
left-- and right handed scalar partner, $\tilde{f}_L$ and $\tilde{f}_{R}$, to
each Standard Model (SM) fermion $f$. These current eigenstates mix to form the
mass eigenstates $\tilde{f}_1$ and $\tilde{f}_2$. The search for these SUSY
scalar fermions is one of the main entries of the LEP2 and Tevatron agendas. At 
the Tevatron, the production cross sections of squarks are rather large since 
they are strongly interacting particles and stringent bounds, $m_{\tilde{q}} 
\gsim 250$ GeV \cite{PDG,Tev}, have been set on the masses of the 
scalar partners of the light quarks by the CDF and D0 collaborations. At LEP2, 
bounds close to the kinematical limits, $m_{\tilde{l}} \gsim 80$ GeV 
\cite{PDG,Lep2}, have been set on the masses of the charged scalar leptons,
while the experimental bound on the mass of the sneutrinos is still rather
low, $m_{\tilde{\nu}} \gsim 45$ GeV \cite{PDG}. \s

The situation of the top squarks is rather special. Indeed, the two current 
stop eigenstates $\tilde{t}_L$ and $\tilde{t}_R$ could strongly mix \cite{qmix}
due to the large $m_t$ value which enters in the non--diagonal element of the 
mass matrix. This leads to a mass eigenstate $\tilde{t}_1$ possibly much 
lighter than the other squarks, and even lighter than the top quark itself. 
If the stop $\tilde{t}_1$ is lighter than the top quark and the chargino [and 
also lighter than the scalar leptons], the two--body decay modes \cite{twobod}
into a top 
quark and the lightest neutralino [which, in the MSSM with conserved R--parity 
\cite{Rp}, is expected to be the lightest SUSY particle (LSP)] and into a 
bottom quark and the lightest chargino, are kinematically forbidden at the 
tree--level. The main $\tilde{t}_1$ decay channel is then expected to be the 
loop--induced and flavor--changing decay into a charm quark and the lightest 
neutralino \cite{Hikasa} 
\begin{eqnarray}
\tilde{t}_1 \ra c\chi_1^0
\end{eqnarray}

At the Tevatron, a light scalar top squark can be produced either directly in 
pairs through gluon--gluon fusion and quark--antiquark annihilation, $gg / q
\bar{q} \ra \tilde{t}_1 \tilde{t}_1^*$ \cite{ppstop}, or in top quark 
decays, $t \ra \tilde{t}_1 \chi_1^0$ \cite{topdec}, if kinematically allowed. 
With the assumption that the branching ratio for the decay $\tilde{t}_1 \ra c
\chi_1^0$ is 100\%, a contour in the $m_{\tilde{t}_1}$--$m_{\chi_1^0}$ plane 
has been excluded by the CDF collaboration in a preliminary analysis 
\cite{CDFstop}. For instance, for a neutralino mass of $m_{\chi_
1^0} \sim 40$ GeV, the maximum excluded $\tilde{t}_1$ mass is $m_{\tilde{t}_1} 
\simeq 120$ GeV; for smaller or larger $m_{\chi_1^0}$ values the bounds on 
$m_{\tilde{t}_1}$ are lower. At LEP2, the lightest top squark is produced in 
pairs through $s$--channel photon and $Z$--boson exchange diagrams, $\ee \ra 
\gamma, Z \ra \tilde{t}_1 \tilde{t}_1^*$ \cite{eestop}. Again, assuming a 
branching ratio of 100\% for the decay $\tilde{t}_1 \ra c\chi_1^0$, the LEP 
collaborations have set a lower bound\footnote{Recently the OPAL collaboration
\cite{Opal} has set a stronger bound of $m_{\tilde{t}_1} \gsim 87.2$ GeV with
a mass splitting between the stop and 
the LSP larger than 10 GeV.} of $m_{\tilde{t}_1} \gsim 
83$ GeV \cite{Lep2} on the lightest stop mass, with the additional assumption 
that the amount of missing energy is larger than 15 GeV \cite{Lep2}.  

\smallskip

All these searches rely on the fact that the decay $\tilde{t}_1 \ra c\chi_1^0$ 
is largely dominant\footnote{Some three body decays of the top squark have
also been considered \cite{porod}. At LEP2 for instance, the possibility of a 
light sneutrino, leading to the kinematically accessible decay mode 
$\tilde{t}_1 \ra b l \tilde{\nu}$, has been analyzed; in this case, a bound 
$m_{\tilde{t}_1} \gsim 85$ GeV \cite{Lep2} has been set on the lightest stop 
mass.}. However, there is another decay mode which is possible in the MSSM, 
even if the lightest top squark is the next--to--lightest SUSY particle: the 
four--body decay into a bottom quark, the LSP and two massless fermions 
\begin{eqnarray}
\tilde{t}_1 \ra b\chi_1^0 f\bar{f}'
\end{eqnarray} 
This decay mode is mediated by virtual top quark, chargino, sbottom, slepton 
and first/second generation squark exchange, Fig.~1, and is of the same order 
of perturbation theory 
as the loop induced  decay $\tilde{t}_1 \ra c \chi_1^0$, i.e. ${\cal O}
(\alpha^3)$. In principle, it can therefore compete with the latter decay
channel. Several estimates of the order of magnitude of the decay rate of the 
process eq.~(2) have been made in the literature \cite{Hikasa,4bodyold}. These 
estimates were based on the assumption of the dominance of one of the 
contributing diagrams [the last diagram of Fig.~1c in Ref.~\cite{Hikasa} 
for instance], and the exchanged particles are assumed to be much heavier 
than the decaying stop squark [therefore working in the point--like limit to 
evaluate the amplitudes]. In this case, 
the output [as one might expect since the virtual 
particles were too heavy from the beginning] was that the decay rate eq.~(2) 
is in general much smaller than the decay rate of the loop induced decay into 
a charm quark and a neutralino. \s

The purpose of this paper is to revisit the four--body decay eq.~(2) in
the light of the recent experimental limits on the masses of the SUSY 
particles. We perform a complete calculation of the decay process, taking 
into account all Feynman diagrams and interference terms. We show that,
if the exchanged particles do not have a too large virtuality [i.e. that
they are not much heavier than the decaying top squark] this four--body 
decay can in fact dominate over the loop induced decay channel eq.~(1) in 
large areas of the MSSM parameter space. This result will therefore affect 
the present experimental lower bounds on the $\tilde{t}_1$ mass, since as 
discussed previously, this state has been searched for under the assumption 
that the decay mode $\tilde{t}_1 \ra c \chi_1^0$ is the main decay channel.
A fortran code calculating the partial widths and branching ratios for 
this four--body  decay mode \cite{fortran} is made available, and the 
lengthy formulae for the four--body decay width will be given elsewhere
\cite{fortran}. \s

The rest of the paper is organized as follows. In section 2, we introduce our 
notation and discuss the two--body decay modes of the top squarks, and in 
particular the decay of the lightest stop squark into charm and neutralino, 
paying special attention to the cases where the decay rate can suppressed. 
In section 3, we analyze the four--body decay mode eq.~(2), and make a 
detailed numerical comparison with the previous decay channel. Some conclusions
are then given in section 4. 

\subsection*{2. The Two--Body Decays} 

In this section, we will first summarize the properties of top squarks: 
masses and mixing, and then discuss their tree--level two--body decays into 
neutralinos and top quarks, and charginos and bottom quarks as well as the 
loop induced decay of the lightest top squark into a charm quark and the 
lightest neutralino.

\subsubsection*{2.1 Squark masses and mixing}

As discussed previously, the left--handed and right--handed squarks of the
third generation $\tilde{f}_L$ and $\tilde{f}_R$ [the current eigenstates] 
can strongly mix to form the mass eigenstates $\tilde{f}_1$ and $\tilde{f}_2$; 
the mass matrices which determine the mixing is given by
\begin{eqnarray}
 M_{\tilde{f}}^2 \ = \ 
\left[ \begin{array}{cc} m_{LL}^2  & m_f \tilde{A}_f  
\\ m_f \tilde{A}_f & m_{RR}^2  
\end{array} \right]
\end{eqnarray}
with, in terms of the soft SUSY--breaking scalar masses $m_{\tilde{f}_L}$ and  
$m_{\tilde{f}_R}$, the trilinear coupling $A_f$, the higgsino
mass parameter $\mu$ and $\tb =v_U/v_D$, the ratio of the vacuum expectation 
values of the two--Higgs doublet fields 
\begin{eqnarray}
m_{LL}^2&=& m_f^2+m_{\tilde{f}_L}^2 + (I_f^3 -e_f s_W^2)\cos 2\beta\,
M_Z^2  \non \\
m_{RR}^2 &=& m_f^2+m_{\tilde{f}_R}^2 + e_f s_W^2\,\cos 2\beta\,M_Z^2 \non \\
\tilde{A}_f &=& A_f-\mu (\tb)^{-2I_f^3}   \hspace*{3cm} 
\end{eqnarray}
with $e_f$ and $I_f^3$ the electric charge and weak isospin of the 
sfermion $\tilde f$ and  $s_W^2=1-c_W^2\equiv \sin^2\theta_W$. 
The mass matrices are diagonalized by $ 2 \times 2$ rotation matrices of 
angle $\theta_f$
\beq
\left(  \begin{array}{c} \tilde f_1 \\ \tilde f_2 \end{array} \right)
={\cal{R}}^{\tilde f} 
\left(  \begin{array}{c} \tilde f_L \\ \tilde f_R \end{array} \right) \non 
\eeq
\beq
 {\cal{R}}^{\tilde{f}} &=&  \left( \begin{array}{cc}
     \ct{f} & \st{f} \\ -\st{f} & \ct{f}
  \end{array} \right)  \ \ \ \ , \ \ \ct{f} \equiv \cos \theta_f 
\ \ {\rm and} \ \ \st{f} \equiv \sin \theta_f 
\eeq
The mixing angle $\theta_f$ and the squark eigenstate masses are then given by 
\begin{eqnarray}
\label{stopmix}
\sin \theta_f = \frac{- m_f \tilde{A}_f} {\sqrt{ (m_{LL}^2-m_{\tilde {f}_1}^2)^2
+m_f^4 \tilde{A}_f^4} } \ \ , \ \ 
\cos \theta_f = \frac{m_{LL}^2 - m_{\tilde{f}_1}^2} 
{\sqrt{ (m_{LL}^2-m_{\tilde {f}_1}^2)^2
+m_f^4 \tilde{A}_f^4}} \hspace*{0.8cm}  \\
m_{\tilde{f}_{1,2}}^2 =  \frac{1}{2} \left[ 
m_{LL}^2 + m_{RR}^2 \mp \sqrt{
(m_{LL}^2 - m_{RR}^2)^2 + 4m_f^2 \tilde{A}_f^2 } \right] \hspace*{1cm}
\end{eqnarray}
In our analysis, we will take into into account the mixing not only in the 
stop sector where it is very important because of the large value of $m_t$, 
but also in the sbottom and stau sectors, where it might be significant for 
large values of the parameters $\mu$ and $\tb$. Furthermore, we will 
concentrate on two scenarii to illustrate our numerical results: \s

$(i)$ ``Unconstrained" MSSM: 
we will assume for simplicity a common soft SUSY--breaking scalar mass for 
the three generations of squarks and sleptons and for isospin up and down 
type particles: $ 
m_{\tilde{t}_L}=m_{\tilde{t}_R}= m_{\tilde{b}_R} \equiv m_{\tilde{q}}$ and
$m_{\tilde{\tau}_L}=m_{\tilde{\tau}_R} \equiv m_{\tilde{l}}$ 
[$m_{\tilde{b}_L}$=$m_{\tilde{t}_L}$ and $m_{\tilde{\nu}_L}$=$m_{\tilde{\tau
}_L}$ by virtue of SU(2) invariance]. The splitting between different particles,
and in particular between the two top squarks, will be then only due to the 
D--terms and to the off--diagonal entries in the sfermion mass matrices. [Note 
that in this case, in most of the parameter space, the stop mixing angle 
is either close to $\pi/2$  (no mixing) or to $\pm \pi/4$ (maximal mixing) 
for respectively, small and large values of the off--diagonal entry 
$m_t \tilde{A}_t$ of the matrix $M_{\tilde{t}}^2$.] Furthermore, we will assume
that the mixing between different generations is absent at the tree level
[otherwise the decay mode $\tilde{t}_1 \ra t \chi_1^0$ will occur already at 
this level]. \s

$(ii)$ Constrained MSSM: or the minimal Supergravity model 
(mSUGRA) where the soft SUSY breaking scalar masses, gaugino 
masses and trilinear couplings are universal at the GUT scale; the left-- 
and right--handed sfermion masses are then given in terms of the gaugino mass 
parameter $m_{1/2}$, the universal scalar mass  $m_0$, the universal
trilinear coupling $A_0$ and $\tb$. 
For the soft SUSY breaking scalar masses, the parameter $\mu$ and the 
trilinear couplings at the low energy scale, we will use the approximate 
formulae for the one--loop  RGEs given in Ref.~\cite{GUTrelations}. 
In mSUGRA [in the small $\tan \beta$ regime], due to the running of the (large) 
top Yukawa coupling, the two stop squarks can be much lighter than the other 
squarks, and in contrast with the first two generations  one has generically a 
sizeable splitting between $m^2_{\tilde{t}_L}$ and $m^2_{\tilde{t}_R}$ at the 
electroweak scale. Thus, even without large mixing $\tilde{t}_1$ can be much 
lighter than the other squarks in this scenario. 

\subsubsection*{2.2 Two--body decays}

If the top squarks are heavy enough, their main decay modes will be 
into top quarks and neutralinos, $\tilde{t}_i \ra t\chi^0_j$ [$j$=1--4],
and bottom quarks and charginos, $\tilde{t}_i \ra b\chi^+_j$ [$j$=1--2]. 
The partial decay widths are given at the tree--level by
\begin{eqnarray}
\Gamma( \tilde{t}_i \ra t \chi_j^0) &=& \frac{\alpha}{4m_{\tilde{t}_i}^3s_W^2}
             \bigg[ ( {a^{\tilde t}_{ij}}^2 + {b^{\tilde t}_{ij}}^2 ) 
                     ( m_{\tilde{t}_i}^2 - m_{t}^2 - m_{\chi_j^0}^2 )
                    - 4 a^{\tilde t}_{ij} b^{\tilde t}_{ij} m_{t} m_{\chi_j^0} \epsilon_{\chi_j}
             \bigg] \non \\ && 
\hspace*{1cm} \lambda^{1/2}(m_{\tilde{t}_i}^2,m_{t}^2,m_{\chi_j^0}^2)   
\nonumber \\
\Gamma(\tilde{t}_i \ra b \chi_j^+) &=& \frac{\alpha}{4 m_{\tilde{t}_i}^3 s_W^2}
             \bigg[ ( {l^{\tilde t}_{ij}}^2 + {k^{\tilde t}_{ij}}^2 )  
                     ( m_{\tilde{t}_i}^2 - m_{b}^2 - m_{\chi_j^+}^2 )
                    - 4 l^{\tilde t}_{ij} k^{\tilde t}_{ij} m_{b} m_{\chi_j^+}
             \bigg] \non \\ &&
\hspace*{1cm} \lambda^{1/2}(m_{\tilde{t}_i}^2,m_{b}^2,m_{\chi_j^+}^2)   
\end{eqnarray}
where $\lambda (x,y,z)=x^2+y^2+z^2-2\,(xy+xz+yz)$ is the usual two--body
phase space function and $\epsilon_{\chi_j}$ is the sign of the eigenvalue
of the neutralino $\chi_j^0$. The couplings $a_{Lj,Rj}^i$ and $b_{Lj,Rj}^i$ 
for the neutral decay are given by 
\begin{eqnarray}
\left\{ \begin{array}{c} a^{\tilde t}_{1j} \\  a^{\tilde t}_{2j} \end{array} \right\}
        &=&   -\frac{m_t N_{j4}}{\sqrt{2} M_W  \sinb}
\left\{ \begin{array}{c} \st{t} \\ \ct{t} \end{array} 
\right\}
         - f_{Lj}\, \left\{ \begin{array}{c} \ct{t} \\ -\st{t} \end{array} 
\right\} \nonumber \\
\left\{ \begin{array}{c} b^{\tilde t}_{1j} \\  b^{\tilde t}_{2j} \end{array} \right\}
        &=&  -\frac{m_t N_{j4}}{\sqrt{2} M_W  \sinb}
\left\{ \begin{array}{c} \ct{t} \\ -\st{t} \end{array} 
\right\}
         - f_{Rj}\, \left\{ \begin{array}{c} \st{t} \\ \ct{t} \end{array} \right\}
\end{eqnarray}
with
\beq
f_{Lj} & = & \sqrt{2}\left[ \frac{2}{3} \; N_{j1}'
              + \left(\frac{1}{2} - \frac{2}{3}\, s_W^2 \right)
                 \frac{1}{c_W s_W}\;N_{j2}' \right]  \nonumber \\ 
f_{Rj} & = &-\sqrt{2}\left[ \frac{2}{3} \; N_{j1}'
              - \frac{2}{3} \frac{s_W}{c_W} \; N_{j2}' \right] \ , 
\eeq
while the couplings $l_{Lj,Rj}^i$ and $k_{Lj,Rj}^i$ for the charged 
decay mode are given by
\beq
\left\{ \begin{array}{c} k^{\tilde t}_{1j} \\ k^{\tilde t}_{2j} \end{array} \right\} & = &
   \frac{m_b\,U_{j2}}{\sqrt{2}\,M_W\,\cosb}
\left\{ \begin{array}{c} \ct{t} \\ -\st{t} \end{array} \right\}
 \non \\
\left\{ \begin{array}{c} l^{\tilde t}_{1j} \\ l^{\tilde t}_{2j} \end{array} \right\} & = &
   V_{j1}\,
\left\{ \begin{array}{c} -\ct{t} \\ \st{t} \end{array} \right\}
 + \frac{m_t\,V_{j2}}{\sqrt{2}\,M_W\,\sinb}
\left\{ \begin{array}{c} \st{t} \\ \ct{t} \end{array} \right\} \ . 
\eeq
In these equations, $N$ and $U/V$ are the diagonalizing matrices for the 
neutralino and chargino states \cite{R10} with 
\beq
N'_{j1}= c_W N_{j1} +s_W N_{j2} \ \ \ , \ \ \
N'_{j2}= -s_W N_{j1} +c_W N_{j2} \ . 
\eeq

\smallskip

If these modes are kinematically not accessible, the lightest top squark  
can decay into a charm quark and the lightest neutralino. This decay mode
is mediated by one--loop diagrams: vertex diagrams as well as squark and quark
self--energy diagrams; bottom squarks, charginos, charged $W$ and Higgs bosons
are running in the loops. The flavor transition $b \ra c$ occurs through the 
charged currents. Adding the various contributions, a divergence is left out
which must be subtracted by adding a counterterm to the scalar self--mass 
diagrams. In Ref.~\cite{Hikasa}, it has been chosen to work in the minimal
Supergravity framework where the squark masses are unified at the GUT scale;
the divergence is then subtracted using a soft--counterterm at $\Lambda_{\rm 
GUT}$, generating a large residual logarithm $\log(\Lambda^2_{\rm GUT}/M_W^2)$  
in the amplitude when the renormalization is performed. This logarithm gives 
the leading contribution to the $\tilde{t}_1 \ra c \chi_1^0$ 
amplitude\footnote{In fact, this contribution is due to the stop--charm mixing 
induced at the one loop level, and which can be obtained by solving the 
renormalization group equations in mSUGRA \cite{Hikasa}.}. \s

Neglecting the non--leading (constant) terms as
well as the charm--quark mass, the partial width of the decay $\tilde{t}_1 
\ra c \chi_1^0$ is given by
\beq
\Gamma( \tilde{t}_1 \ra c \chi_j^0) = \frac{\alpha}{4} m_{\tilde{t}_1} \
\left(1- \frac{m_{\chi_1^0}^2} {m_{\tilde{t}_1}^2 } \right)^2 |f_{L1}|^2 \ 
\epsilon
\eeq
where $f_{L1}$ is given by
\beq
f_{L1} & = & \sqrt{2}\left[ \frac{2}{3} (c_W N_{11} +s_W N_{12})
              + \left(\frac{1}{2} - \frac{2}{3}\, s_W^2 \right)
                 \frac{-s_W N_{11} +c_W N_{12}}{c_W s_W}\right] 
\eeq
and $\epsilon$ denotes the amount of 
$\tilde{t}_{L,R}$--$\tilde{c}_L$ mixing and is given by \cite{Hikasa}:
\begin{eqnarray}
\epsilon = \frac{ \Delta_L c_{\theta_t} - \Delta_R s_{\theta_t} }{ 
m_{\tilde{c}_L}^2 - m_{\tilde{t}_1}^2 }
\end{eqnarray}
with
\begin{eqnarray}
\Delta_L &=& \frac{\alpha}{4 \pi s_W^2} \log \left(\frac{\Lambda_{\rm GUT}^2}
{M_W^2} \right)  \, \frac{V^*_{tb} V_{cb} \, m_b^2}{2M_W^2 \cos^2 \beta} 
\bigg(m_{\tilde{c}_L}^2+m_{\tilde{b}_R}^2 + m_{H_1}^2 +A_b^2\bigg) \non \\
\Delta_R &=&  \frac{\alpha}{4 \pi s_W^2} \log \left(\frac{\Lambda_{\rm GUT}^2}
{M_W^2} \right) \, \frac{V^*_{tb} V_{cb} \, m_b^2}{2M_W^2 \cos^2 \beta}m_t A_b  
\end{eqnarray}
Assuming proper Electroweak symmetry breaking, the Higgs scalar mass $m_{H_1}$ 
can be written in terms of $\mu, \tb$ and the pseudoscalar Higgs boson mass 
$M_A$ as
\beq
m_{H_1}^2= M_A^2 \sin^2\beta - \cos 2\beta M_W^2 -\mu^2
\eeq
Note also that $\Delta_L$ and $\Delta_R$ are suppressed by the CKM matrix
element $V_{cb} \sim 0.05$ and the (running) $b$ quark mass squared $m_b^2
\sim (3$ GeV$)^2$, but very strongly enhanced by the term $\log \left(
\Lambda_{\rm GUT}^2/ M_W^2 \right)$ which is close to $\sim 65$ for
$\Lambda_{\rm GUT} \simeq 2 \cdot 10^{16}$ GeV. 

\bigskip

Let us now discuss the various scenarii in which the decay rate eq.~(13) is 
small. \s
 
$(i)$ First, and as discussed previously, the large logarithm $\log \left(
\Lambda_{\rm GUT}^2/ M_W^2 \right) \sim 65$ appears only because the  
choice of the renormalization condition is made at the GUT scale. This 
might be justified in the framework of the mSUGRA model, but in a general 
MSSM where the squark masses are not unified at some very high scale  
such as $\Lambda_{\rm GUT}$, one could chose a low energy counterterm; in 
this case no large logarithm would appear. In fact, one could have simply
made the renormalization in the $\overline{\rm MS}$ (or $\overline{\rm DR}$)
scheme, where the divergence is simply subtracted, and we would have
been left only with the (very small) subleading terms. \s

$(ii)$ If the lightest top squark is a pure right--handed state [a situation 
which is in fact favored by the stringent constrains \cite{leprho} from 
high--precision electroweak data, and in particular from the $\rho$ parameter; 
see Ref.~\cite{drho} for instance], there is no mixing in the stop sector, 
and the $\epsilon$ term in eq.~(15) involves only the $\Delta_R$ component. 
For moderate values of the trilinear coupling $A_b$, this component can be 
made small enough to suppress the decay rate $\Gamma(\tilde{t}_1 \ra c 
\chi_1^0)$. In addition, the charm squark mass can be made different from 
the lightest stop mass, and taken to be very large; there will be then a 
further strong suppression from the denominator of eq.~(15) [this situation 
occurs in fact also in the mSUGRA model, since because of the running of the 
top Yukawa coupling, $\tilde{t}_R$ can be much lighter than $\tilde{c}_L$, 
especially for large values of the parameter $m_0$; see 
Ref.~\cite{GUTrelations}]. \s

$(iii)$ Even in the case of mixing, for a given choice of the MSSM parameters, 
large cancellations can occur between the various terms in the numerator of 
eq.~(15). Indeed, for some values of the soft SUSY--breaking scalar masses 
and trilinear coupling $A_b$, the two terms $\Delta_{L,R}$ weighted by the 
sine and cosine of the mixing angle might cancel each other; this would happen 
for a value of $\theta_t$ such that $\tan \theta_t\simeq \Delta_L/\Delta_R$. In 
addition, the coefficient $f_{L}$ in eq.~(14), which summarizes the 
gaugino--higgsino texture of the lightest neutralino might also be very small. 
In fact, as can be seen, the parameter involves only the gaugino components 
$N_{11}$ and $N_{12}$; if the neutralino $\chi_1^0$ is higgsino--like, these 
two components are very small, leading to a very small value for $f_{L}$. \s

Thus, there are many situations in which the decay rate $\Gamma (\tilde{t}_1
\ra c \chi_1^0)$ might be very small, opening the possibility for the 
four--body decay mode, to which we turn our attention now, to dominate.  

\subsection*{3. The Four--Body decay mode}

\subsubsection*{3.1 Analytical Results}

The four--body decay mode $\tilde{t}_1 \ra b \chi_1^0 f \bar{f}'$, which 
occurs for stop masses larger than $m_b+m_{\chi_0^1}$, proceeds through
several diagrams as shown in Fig.~1. There are first the $W$--boson exchange 
diagrams with a virtual top quark, bottom squark or the two charginos states
[Fig.~1a]. A similar set of diagrams is obtained by replacing the $W$--boson 
by the charged Higgs boson $H^+$ [Fig.~1b]. A third type of diagrams consist 
of up and down type slepton or first/second generation squark exchanges  
[Fig.~1c]. \s

We have calculated the amplitude squared of the decay mode, taking into 
account all these diagrams and interferences. The complete expressions are
too lengthy and will be given elsewhere \cite{fortran}.
We have taken into account the $b$--quark mass
[which might be important for nearly degenerate stop and LSP masses] and
the full mixing in the third generation sector.  We have then integrated
the amplitude squared over the four--body phase space using the Monte--Carlo
routine Rambo \cite{Rambo}, to obtain the partial decay width. Let us 
summarize the main features of the result. \s

\nn \underline{The charged Higgs boson exchange diagrams Fig.~1b:} \\
These diagrams do not give rise 
to large contributions for two reasons. First, because of the relation
in the MSSM between the charged and pseudoscalar Higgs boson masses, $M_{H^\pm}
^2= M_A^2+M_W^2$, and the experimental bound $M_A \gsim 90$ GeV, the charged 
Higgs boson with mass a $M_{H^\pm} \gsim 120$ GeV has a much larger virtuality 
than the $W$ boson contribution; since the other exchanged particles are
the same, the $H^\pm$ contribution is much smaller than the $W$ contribution.
In addition, the contributions are suppressed by the very tiny Yukawa couplings
of the $H^\pm$ bosons to leptons and light quarks, except in the case of the 
$H^+\nu \tau$ coupling which can be enhanced for large $\tb$ values; however, 
in this case the decay width $\Gamma(\tilde{t}_1 \ra c \chi_1^0)$ [which
grows as $1/\cos^2\beta]$ is also strongly enhanced. Therefore, these 
contributions can be safely neglected in most of the parameter space. \s

\nn \underline{The squark exchange diagrams in Fig.~1a and Fig.~1c:}\\
In general, the diagrams of Fig.~1c  give very small 
contributions for $\tilde{t}_1$ masses of the order of 100 GeV, since the 
first and second generation squarks are expected to be much heavier, $m_{
\tilde{q}} \gsim 250$ GeV, and their virtuality is therefore too large. For 
much larger $\tilde{t}_1$ masses, the two--body decay channel $\tilde{t}_1 \ra
b \chi_1^+$ [or one of the three body decay modes] will be in general open 
and will largely dominate the rate. 
The sbottom contribution in Fig.~1a is also very small, if the mixing 
in the sbottom sector is neglected and the Tevatron bound $m_{\tilde{b}} \gsim 
250$ GeV applies [this bound is valid only if bottom squarks are approximately
degenerate with first/second generation squarks; in the general case the 
experimental bound \cite{Lep2} is lower]. For large values of $\tb$ which 
would lead to a strong 
mixing in the sbottom sector with a rather light $\tilde{b}_1$, the decay
width $\tilde{t}_1 \ra c \chi_1^0$ becomes very large as discussed previously, 
leaving little chance to the four--body decay mode to occur. \s

\nn \underline{Top quark exchange diagrams Fig.~1a:}\\
The contribution of the diagram with an exchanged top quark is only important
if the stop mass is of the order of $m_t+m_{\chi_1^0}$ and therefore 
$m_{\tilde{t}_1} \gsim {\cal O}(250$ GeV), with the proviso that: (i) the 
lightest chargino must be heavier than $\tilde{t}_1$ for the two--body decay 
mode $\tilde{t}_1 \ra b \chi_1^+$ to be kinematically forbidden, and (ii) the 
$\tilde{t}_1$ mass must be smaller than  $m_{\chi_1^0}+M_W$ to forbid the 
three body decay $\tilde{t}_1 \ra b \chi^0_1 W$. In models with gaugino mass
unification, $M_2 \sim M_1/2$, the  $\chi_1^0$ and $\chi_1^+$ masses are related
in such a way that the above conditions [with a not too virtual top quark] are 
fullfiled only in a marginal area of the parameter space. This contribution
can be thus also neglected. \s

\nn \underline{Slepton exchange diagrams Fig.~1c}: \\
In contrast to squarks, slepton [and especially sneutrino] exchange 
diagrams might give substantial contributions, since sleptons masses of ${\cal 
O}$(100 GeV) are still experimentally allowed. In fact, when the difference 
between the lightest stop, the lightest chargino and the slepton masses is 
not large, the diagrams Fig.~1c will give the dominant contribution to the 
four--body decay mode, with a rate possibly much larger than the rate for the 
loop induced decay $\tilde{t}_1 \ra c \chi_1^0$, for small enough values of 
$\tb$. \s

\nn \underline{Chargino exchange diagram Fig.~1a}:\\
The most significant contributions to the four--body decay mode will 
come in general from this diagram, when the virtuality
of the chargino is not too large. In particular, for an exchanged $\chi_1^+$
with a mass not much larger than the experimental lower bound $m_{\chi_1^+} 
\gsim 95$ GeV [but with $m_{\chi_1^+}<m_{\chi_1^0}+M_W$ to forbid the 
three--body decay mode], the decay width can 
be substantial even for top squark masses of the order of 80 GeV. \s

Thus, a good approximation [especially for a light top squark $m_{\tilde{t}_1}
\sim {\cal O}(100$ GeV)] is to take into account only the lightest chargino and 
slepton exchange contributions. 
In terms of the momenta of the various particles involved in the process
[all momenta are outgoing, except for the momentum of the decaying stop which 
is ingoing] and defining the propagators as $D_X=p_X^2-M_X^2$, the amplitude
squared for the charged slepton and sneutrino contributions is given by
[note that $m_{\chi_1^0}$ is the physical LSP mass and $\epsilon_{\chi_1}$
is the sign of the eigenvalue of the neutralino $\chi_1^0$]
\beq 
|A_{\tilde l}|^2 && = \frac{4e^6}{s_W^6 D_{\chi^+_1}^2 D_{\tilde l_L}^2}
m_{\chi^+_1}^2 a_{l}^2 (U_{11})^2 (l_{11}^{\tilde t})^2
(p_{l}.p_{\chi^0_1}) (p_b.p_{\nu}) \\
&& + \frac{4e^6}{s_W^6 D_{\chi_1^+}^2D_{\tilde \nu_L}^2}
a_{\nu }^2(V_{11} )^2 (l^{\tilde t}_{11})^2
(p_\nu.p_{\chi^0_1} )[2(p_b.p_{\chi^+_1} )(p_l.p_{\chi^+_1})-(p_{\chi^+_1})^2
(p_b.p_l)] \non \\
&&+ \frac{4e^6 (l^{\tilde t}_{11})^2
U_{11} V_{11} a_{\nu} a_{l} }
{s_W^6 D_{\tilde \nu_L}D_{\tilde l_L} D^2_{\chi_1^+}}
\epsilon_{\chi_1} m_{\chi^0_1}m_{\chi^+_1} 
\left[ (p_b.p_{\nu})(p_{l}.p_{\chi^+_1})-(p_b.p_{l})(p_{\nu}.p_{\chi^+_1})
+(p_b.p_{\chi^+_1})(p_{\nu}.p_{l})\right] \non 
\eeq
where we have taken into account only the exchange of the lightest chargino,
and neglected the mixing in the charged slepton sector [which is in general
not too large for small values of $\tb$]. In terms of the
elements of the matrices  diagonalizing the chargino and neutralino mass 
matrices, the couplings $a_f$ and $l^{\tilde t}_{11}$ read
\begin{eqnarray}
l^{\tilde t}_{11}=-c_{\theta_t} V_{11} +s_{\theta_t} Y_t V_{12} \ \ , \ \
Y_{t} = \frac{m_{t}}{\sqrt{2} M_W \sin \beta} \hspace*{0.2cm} \non  \\
a_f = -\sqrt{2} e_f N_{11} s_W 
-\sqrt{2}(I_3^f-e_fs_W^2) \frac{N_{12}}{c_W} 
\eeq
Taking into account only the contribution of the lightest chargino, 
the amplitude squared for the chargino exchange diagram is given by:
\beq
|A_{\chi^+_1}|^2 &&= \frac{8N_ce^6}{s_W^6 D_W^2 D_{\chi^+_1}^2} 
(l^{\tilde t}_{11})^2 \bigg[ m_{\chi^+_1}^2 (O^R_{11})^2 (p_b.p_{f})
(p_{\chi^0_1}.p_{f'}) \non \\
&&-\epsilon_{\chi_1}m_{\chi^0_1}m_{\chi^+_1}O^L_{11}O^R_{11}
\bigg( (p_{f}.p_{f'})(p_b.p_{\chi^+_1})-(p_{f}.p_{\chi^+_1})(p_b.p_{f'})
+(p_b.p_{f})(p_{\chi^+_1}.p_{f'}) \bigg)  \non \\
&&+ (O^L_{11})^2 \bigg( 2(p_{\chi^0_1}.p_{f}) (p_{\chi^+_1}.p_{f'})(p_{\chi^+_1}.p_b)
-p_{\chi^+_1}^2(p_{\chi^0_1}.p_{f})(p_b.p_{f'}) \bigg) \bigg]
\eeq
with $N_c$ the color factor, and the chargino--neutralino--W couplings
$O^{L,R}_{11}$ [the coupling $l^{\tilde t}_{11}$ was given above]
\beq
O^L_{11}=-\frac{V_{12}}{\sqrt{2}}N_{14}+V_{11}N_{12} \ \ , \ \ 
O^R_{11}=\frac{U_{12}}{\sqrt{2}}N_{13}+U_{11}N_{12}
\eeq
If the $W$ boson in the virtual chargino diagram decays leptonically, one
has also to take into account the interference between the chargino and
slepton exchange diagrams. Assuming again no mixing in the charged slepton 
sector [which means that only left--handed sleptons contribute], the 
interference term is given by: 
\begin{eqnarray}
Re[A_{\chi^+_1}A^*_{\tilde l}] &&=
\nonumber
\frac{- 4e^6(l^{\tilde t}_{11})^2V_{11}a_{\nu}}
{s_W^6\sqrt{2}D_{\chi_1^+}^2D_WD_{\tilde \nu_L}}
[ 2 O^L_{11} \left[2(p_b.p_{\chi^+_1}) (p_{\chi^+_1}.p_{l}) (p_{\nu}.p_{\chi^0_1})
- (p_b.p_{l}) p_{\chi^+_1}^2 (p_{\nu}.p_{\chi^0_1})\right]
\\ \nonumber
&& 
+\epsilon_{\chi_1}
m_{\chi^+_1} m_{\chi^0_1} O^R_{11}\left[(p_b.p_l) (p_{\chi^+_1}.p_{\nu})
-(p_b.p_{\nu}) (p_{\chi^+_1}.p_{l})-(p_b.p_{\chi^+_1}) (p_{\nu}.p_{l}) \right]]
\\ 
&&
+\frac{4e^6(l^{\tilde t}_{11})^2a_l
U_{11}}{s_W^6\sqrt{2}D_{\chi^+_1}^2
D_WD_{\tilde l_L}} m_{\chi^+_1}
[-2m_{\chi^+_1}O^R_{11}(p_b.p_{\nu})(p_{\chi^0_1}.p_{l})
\\ \nonumber
&&
+\epsilon_{\chi_1}
m_{\chi^0_1}O^L_{11}
[(p_b.p_{\nu})(p_{\chi^+_1}.p_{l})
+(p_b.p_{\chi^+_1})(p_{\nu}.p_{l})-(p_b.p_{l})(p_{\chi^+_1}.p_{\nu})]
\end{eqnarray}

Note that if the exchanged $W$ bosons and sleptons are real [i.e. for 
$\tilde{t}_1$ masses larger than $M_W+m_{\chi_1^0}+m_b$ and/or $m_{\tilde{l}}
+m_b$ respectively], the three body decay channels $\tilde{t}_1 \ra 
b W^+ \chi_1^0$ and $\tilde{t}_1 \ra b l \tilde{l}'$ open up. This situation 
can be handled by including the total widths of the $W$ boson and sleptons 
in their respective propagators. In the numerical analysis though, we will 
concentrate only on the kinematical regions where the $W$ boson and sleptons 
are off mass--shell.

\subsubsection*{3.2 Numerical Results}

For the numerical results, where we include the contributions of all diagrams
[and not only the dominant chargino and slepton exchange diagrams discussed 
above] we first show in Figs.~2 and 3, the branching ratios for the four--body 
decay mode BR($\tilde{t}_1 \ra b\chi_1^0 f \bar{f}')$ in the unconstrained 
MSSM scenario, with a common squark mass $m_{\tilde{q}}$. 
For the gaugino sector, we have chosen $\tb=2.5$, a gaugino mass parameter 
$M_2=120$ GeV [Figs.~2a and 3a] and 200 GeV [Figs.~2b and 3b] and three values 
of the parameter $\mu=300, 450$ and 700 GeV. This leads to the lightest 
chargino and neutralino masses shown in Table 1; these values allow for 
charginos $\chi_1^+$ with a not too large virtuality, but still heavy 
enough to comply with the available experimental bounds [even for the choice 
$M_2=120$ GeV]. 
The trilinear coupling $A_t$ is varied to fix $m_{\tilde{t}_1}$
to a constant value, while the coupling $A_b$ is fixed to $A_b=-100$ GeV. 
For the additional parameters which enter the $\tilde{t}_1 \ra c \chi_1^0$ 
amplitude, we will take [for the entire numerical analysis, i.e. also for the 
other figures]: $M_A=500$ GeV [except in the mSUGRA scenario discussed later,
where $M_A$ is given by the RGE's], $V_{cb}=0.05$, $m_b^{\rm run.}=3$ GeV; the 
cut--off $\Lambda$ will be taken to be the GUT scale $\Lambda_{\rm GUT}=2
\cdot 10^{16}$ GeV. 
\bigskip 

\begin{table}[hbt]
\renewcommand{\arraystretch}{1.5}
\begin{center}
\begin{tabular}{|c||c||c|c|c|} \hline
\ \ $m_{\tilde{t}_1}$ \ \ & \ \ $M_2$ \ \ & \ \ $\mu$ \ \ & \ \ 
$m_{\chi_1^0}$ \ \ & \ \ $m_{\chi_1^+}$ \ \ \\ 
\hline \hline
80 & 120 & 300 & 50.2 & 96 \\
   &     & 450 & 53.6 & 106 \\
   &     & 700 & 55.4 & 112 \\ \hline 
150 & 200 & 300 & 87.8 & 163 \\
    &     & 450 & 91.8 & 181 \\
    &     & 700 & 93.8 & 190 \\ \hline 
\end{tabular}
\end{center}
\caption[]{$\chi_1^0$ and $\chi_1^+$ masses for $\tb=2.5$ and the choices of 
$M_2$ and $\mu$ parameters of Figs.~2--3. All parameters and masses are in GeV.}
\end{table}

In Fig.~2, the branching ratio BR$(\tilde{t}_1 \ra b\chi_1^0 f \bar{f}')$ is
shown as a function of the soft SUSY breaking scalar mass $m_{\tilde{q}}$ for 
two values of the lighter top squark mass $m_{\tilde{t}_1}=80$ GeV [2a] and 150
GeV [2b]. The common slepton mass is taken to be $m_{\tilde{l}} \sim 200$ GeV 
[so that the contribution of the sleptons to the decay will be extremely small
especially for the small $m_{\tilde{t}_1}$ values].  
One sees that even for a rather light stop, $m_{\tilde{t}_1}=80$ GeV [Fig.~2a], 
the branching ratio can be largely dominating.
For $\mu=300$ GeV, it is already the case for values of $m_{\tilde{q}}$ 
close to $\sim 200$ GeV [which is needed to keep the masses of the first and
second generation squarks larger than the present experimental lower bound]; 
for larger $m_{\tilde{q}}$ values, the branching ratio becomes very close to 
one. For the value $\mu=450$ GeV, the branching ratio drops to the level of 10 
to 20\%; this is mainly due to the positive interference generated by the soft 
scalar mass squared $m_{H_1}^2 \sim - \mu^2$ [eq.~(17)] which enhances the 
decay rate $\Gamma(\tilde{t}_1 \ra c \chi_1^0$), but also to the fact that 
the exchanged chargino [which gives the largest contribution to the four--body 
decay channel] is heavier than in the previous case, so that its large
virtuality suppresses 
the four--body decay mode. For the value $\mu=700$ GeV, the 
interference can be either positive or negative, and in the margin  $m_{\tilde{
q}} \sim $ 300--400 GeV, the decay $\tilde{t}_1 \ra b\chi_1^0 f \bar{f}'$ 
dominates over the loop induced decay $\tilde{t}_1 \ra c\chi_1^0$ in spite of 
the even larger virtuality  of the chargino. \s

For a heavier $\tilde{t}_1$ state, $m_{\tilde{t}_1} =150$ GeV [Fig.~2b], and 
for the value $M_2=200$ GeV [leading to heavier charginos and neutralinos] and
the same $\mu$ values as above, the branching ratio BR$(\tilde{t}_1 \ra b 
\chi_1^0 f \bar{f}')$ is larger than 95\% for almost all values of the 
parameter $m_{\tilde{q}}$. This is first due to the fact that the phase space 
is more favorable in this case [i.e. the virtuality of the chargino is 
relatively smaller], and also to the fact that for this particular choice
of the lightest stop mass, the parameter $\epsilon$ in eq.~(15) which governs
the magnitude of the decay rate  $\Gamma(\tilde{t}_1 \ra c\chi_1^0$) is
suppressed [i.e. the mixing angle is such that, there is a partial cancellation
between the two terms in the numerator of eq.~(15)]. \s

Figs.~3 show the branching ratio BR$(\tilde{t}_1 \ra b\chi_1^0 f \bar{f}')$
as a function of the lightest stop mass, for a fixed value of the common
squark and slepton mass $m_{\tilde{q}}=m_{\tilde{l}}=400$ GeV, and for 
the same parameters $\tb$, $\mu$ and $M_2$ as in Fig.~2. As can be seen,
the branching ratio is very small in the lower stop mass range where the 
virtuality of the exchanged chargino is rather large, and increases with 
increasing $m_{\tilde{t}_1}$ to reach values close to unity near the 
$m_{\tilde{t}_1} \sim m_b +m_{\chi_1^+}$ threshold where the two--body decay 
mode $\tilde{t}_1 \ra b \chi_1^+$ opens up [of course, beyond this threshold, 
the loop induced decay 
becomes irrelevant and we stopped the curves at these values]. Note that even 
for the small values $m_{\tilde{t}_1} \sim 80$ GeV [Fig.~3a], the branching 
ratio for the four--body decay mode $\tilde{t}_1\ra b\chi_1^0f\bar{f}'$ can 
reach the level of 90\%. Since top squarks with these mass values have been 
experimentally ruled out by LEP2 [and possibly Tevatron] searches under the 
assumption that they decay most of time into charm quarks and the lightest 
neutralinos, the searches at LEP2 have to be reconsidered in the light of the 
possible dominance of the four--body decay mode. \s 

In the previous figures, sleptons were too heavy to contribute substantially
to the decay rate  $\tilde{t}_1\ra b\chi_1^0f\bar{f}'$ since we assumed 
the common slepton mass to be $m_{\tilde{l}} = m_{\tilde{q}} >200$ GeV. In 
Fig.~4, we relax the assumption $m_{\tilde{l}} = m_{\tilde{q}}$ and show the 
branching ratio BR($\tilde{t}_1\ra b\chi_1^0f\bar{f}')$ as a function of
the sneutrino mass $m_{\tilde{\nu}}$ [the masses of the other slepton are
then fixed and are of the same order] for $m_{\tilde{t}_1}=80$ GeV, $\mu=300$ 
GeV [Fig.~4a] and 700 GeV [Fig.~4b] and three values of the common soft scalar 
quark mass $m_{\tilde{q}}=300,500$ and 800 GeV. As can be seen, the 
contribution of sleptons can substantially enhance the four--body decay 
branching for relatively small masses [corresponding to $m_{\tilde{\nu}} \lsim 
120$ GeV in this case]. For larger sneutrino masses, the sleptons become too
virtual and we are left only with the contribution of the lightest chargino 
discussed previously [and which is constant in this case]. \s

Finally, let us turn to the case of the mSUGRA scenario. The branching ratio 
of the decay $\tilde{t}_1\ra b \chi_1^0f\bar{f}'$ for a light top squark 
$m_{\tilde{t}_1} \sim$ 70--130 GeV is shown in Fig.~5, as a function of 
$m_{\tilde{t}_1}$ for $\tb=2.5$, $\mu>0$ [Fig.~5a] and $\mu<0$ [Fig.~5b] and 
several choices of the value of the gaugino mass parameter $m_{1/2}\sim 0.8 
M_2$ [again, the choice of $m_{1/2}$ leads to chargino masses not too much 
larger than the allowed experimental bounds, $m_{\chi_1^+} \lsim 140$ GeV; in 
fact in this scenario, $\mu$ is always large and the neutralinos and charginos 
and almost bino and wino like, with masses $m_{\chi_1^0} \sim M_2/2$ and 
$m_{\chi_1^+} \sim M_2$]. Again, one sees that BR($\tilde{t}_1\ra b \chi_1^0f
\bar{f}'$) can be very large, exceeding in some cases the 50\% level, even for 
values $m_{\tilde{t}_1} \sim 80$ GeV, which are  experimentally 
excluded by the negative search of the $\tilde{t}_1 \ra c \chi_1^0$ signature
{\em if} this decay channel dominates.  
  
\subsection*{4. Conclusions} 

We have analyzed the four--body decay mode of the lightest top squark into
the lightest neutralino, a bottom quark and two massless fermions, $\tilde{t}_1
\ra b \chi_1^0 f \bar{f}'$, in the framework of the minimal supersymmetric 
extention of the Standard Model, where the neutralino $\chi_1^0$ is expected 
to be the lightest SUSY particle. Although we have evaluated the partial 
decay width taking into account all the contributing diagrams [and their 
interferences], we have singled out those which give the dominant 
contributions. \s

For small $\tilde{t}_1$ masses accessible at LEP2 and the Tevatron, we have
shown that this four--body decay mode can dominate over the loop--induced 
decay into a charm quark and the LSP, $\tilde{t}_1 \ra c \chi_1^0$, if 
charginos and sleptons have masses not too much larger than their present
experimental bounds. This holds in the case of both the ``unconstrained" and 
constrained (mSUGRA) MSSM. \s

This result will affect the experimental searches of the lightest top squark 
at LEP2 and at the Tevatron, since only the charm plus lightest neutralino 
signal has been considered so far in these experiments. However, the topology of
the four--body decay is similar to the ones of the three body decay mode 
$\tilde{t}_1 \ra b l \tilde{l}'$ [for final state leptons] which has been 
searched for at LEP2 \cite{Lep2} and of the two--body decay mode $\tilde{t}_1
\ra b \chi_1^+$ which has been looked for at the Tevatron \cite{CDFstop}. 
The extension of the experimental searches to the decay mode 
$\tilde{t}_1 \ra b \chi_1^0 f \bar{f}'$ should be thus straightforward. 

\vspace*{1cm}

\nn {\bf Acknowledgements:} \\
We thank Manuel Drees, Wolfgang Hollik and Gilbert Moultaka for discussions.  
This work is supported by the french ``GDR--Supersym\'etrie".

\newpage

\begin{picture}(1000,1200)(0,0)
\Text(200,630)[]{Figure 1: Feynman diagrams contributing to the four--body decay mode $ \tilde{t}_1 \ra b\chi_1^0 f\bar{f}'$.}
\Text(0,1200)[]{{\bf a)}}
\Text(0,1000)[]{{\bf b)}}
\Text(0,800)[]{{\bf c)}}
\Text(0,1120)[]{$\tilde{t}$}
\DashArrowLine(10,1120)(40,1120){4}{}
\ArrowLine(70,1150)(40,1120)
\Text(77,1150)[]{$\chi_0$}
\ArrowLine(40,1120)(60,1090)
\Text(35,1100)[]{$t$}
\ArrowLine(60,1090)(90,1120)
\Text(100,1120)[]{$b$}
\Photon(80,1065)(60,1090){4}{8}
\Text(55,1075)[]{$W$}
\ArrowLine(80,1065)(110,1095)
\Text(115,1090)[]{$f$}
\ArrowLine(110,1035)(80,1065)
\Text(115,1035)[]{$\overline{f}'$}


\Text(150,1120)[]{$\tilde{t}$}
\DashArrowLine(160,1120)(190,1120){4}{}
\Photon(190,1120)(220,1150){4}{8}
\Text(200,1150)[]{$W$}
\ArrowLine(220,1150)(250,1180)
\Text(255,1180)[]{$f$}
\ArrowLine(250,1150)(220,1150)
\Text(255,1150)[]{$\overline{f}'$}
\DashArrowLine(210,1090)(190,1120){4}{}
\Text(187,1100)[]{$\tilde{b}$}
\ArrowLine(240,1120)(210,1090)
\Text(250,1120)[]{$\chi_0$}
\ArrowLine(210,1090)(240,1060)
\Text(235,1079)[]{$b$}


\Text(300,1120)[]{$\tilde{t}$}
\DashArrowLine(310,1120)(340,1120){4}{}
\ArrowLine(340,1120)(370,1150)
\Text(377,1150)[]{$b$}
\ArrowLine(340,1120)(360,1090)
\Text(335,1100)[]{$\chi^+_i$}
\ArrowLine(390,1120)(360,1090)
\Text(400,1120)[]{$\chi_0$}
\Photon(380,1065)(360,1090){4}{8}
\Text(360,1075)[]{$W$}
\ArrowLine(380,1065)(410,1095)
\Text(415,1090)[]{$f$}
\ArrowLine(410,1035)(380,1065)
\Text(415,1035)[]{$\overline{f}'$}

\vspace*{-1.5cm}


\Text(0,920)[]{$\tilde{t}$}
\DashArrowLine(10,920)(40,920){4}{}
\ArrowLine(70,950)(40,920)
\Text(77,950)[]{$\chi_0$}
\ArrowLine(40,920)(60,890)
\Text(35,900)[]{$t$}
\ArrowLine(60,890)(90,920)
\Text(100,920)[]{$b$}
\DashArrowLine(60,890)(80,865){4}{}
\Text(55,875)[]{$H^+$}
\ArrowLine(80,865)(110,895)
\Text(115,890)[]{$f$}
\ArrowLine(110,835)(80,865)
\Text(115,835)[]{$\overline{f}'$}


\Text(150,920)[]{$\tilde{t}$}
\DashArrowLine(160,920)(190,920){4}{}
\DashArrowLine(190,920)(220,950){4}{}
\Text(200,950)[]{$H^+$}
\ArrowLine(220,950)(250,980)
\Text(255,980)[]{$f$}
\ArrowLine(250,950)(220,950)
\Text(255,950)[]{$\overline{f}'$}
\DashArrowLine(210,890)(190,920){4}{}
\Text(185,900)[]{$\tilde{b}$}
\ArrowLine(240,920)(210,890)
\Text(250,920)[]{$\chi_0$}
\ArrowLine(210,890)(240,860)
\Text(237,879)[]{$b$}


\Text(300,920)[]{$\tilde{t}$}
\DashArrowLine(310,920)(340,920){4}{}
\ArrowLine(340,920)(370,950)
\Text(377,950)[]{$b$}
\ArrowLine(340,920)(360,890)
\Text(335,900)[]{$\chi_i^+$}
\ArrowLine(390,920)(360,890)
\Text(400,920)[]{$\chi_0$}
\DashArrowLine(360,890)(380,865){4}{}
\Text(360,875)[]{$H^+$}
\ArrowLine(380,865)(410,895)
\Text(415,890)[]{$f$}
\ArrowLine(410,835)(380,865)
\Text(415,835)[]{$\overline{f}'$}

\vspace*{-.8cm}

\Text(50,740)[]{$\tilde{t}$}
\DashArrowLine(60,740)(90,740){4}{}
\ArrowLine(90,740)(120,770)
\Text(127,770)[]{$b$}
\ArrowLine(90,740)(110,710)
\Text(85,720)[]{$\chi_i^+$}
\ArrowLine(140,740)(110,710)
\Text(150,740)[]{$\overline{f}'$}
\DashArrowLine(110,710)(130,685){4}{}
\Text(110,695)[]{$\tilde{f}$}
\ArrowLine(130,685)(160,715)
\Text(165,710)[]{$f$}
\ArrowLine(160,655)(130,685)
\Text(170,655)[]{$\chi_0$}


\Text(250,740)[]{$\tilde{t}$}
\DashArrowLine(260,740)(290,740){4}{}
\ArrowLine(290,740)(320,770)
\Text(327,770)[]{$b$}
\ArrowLine(290,740)(310,710)
\Text(285,720)[]{$\chi_i^+$}
\ArrowLine(310,710)(340,740)
\Text(350,740)[]{$f $}
\DashArrowLine(330,685)(310,710){4}{}
\Text(310,685)[]{$\tilde{f^*}'$}
\ArrowLine(360,715)(330,685)
\Text(365,710)[]{$\overline{f}'$}
\ArrowLine(330,685)(360,655)
\Text(370,655)[]{$\chi_0$}
\end{picture}

\newpage
\setcounter{figure}{1}
\begin{figure}[htb]
\vspace*{-3.5cm}
\hspace*{-2cm}
\mbox{
\psfig{figure=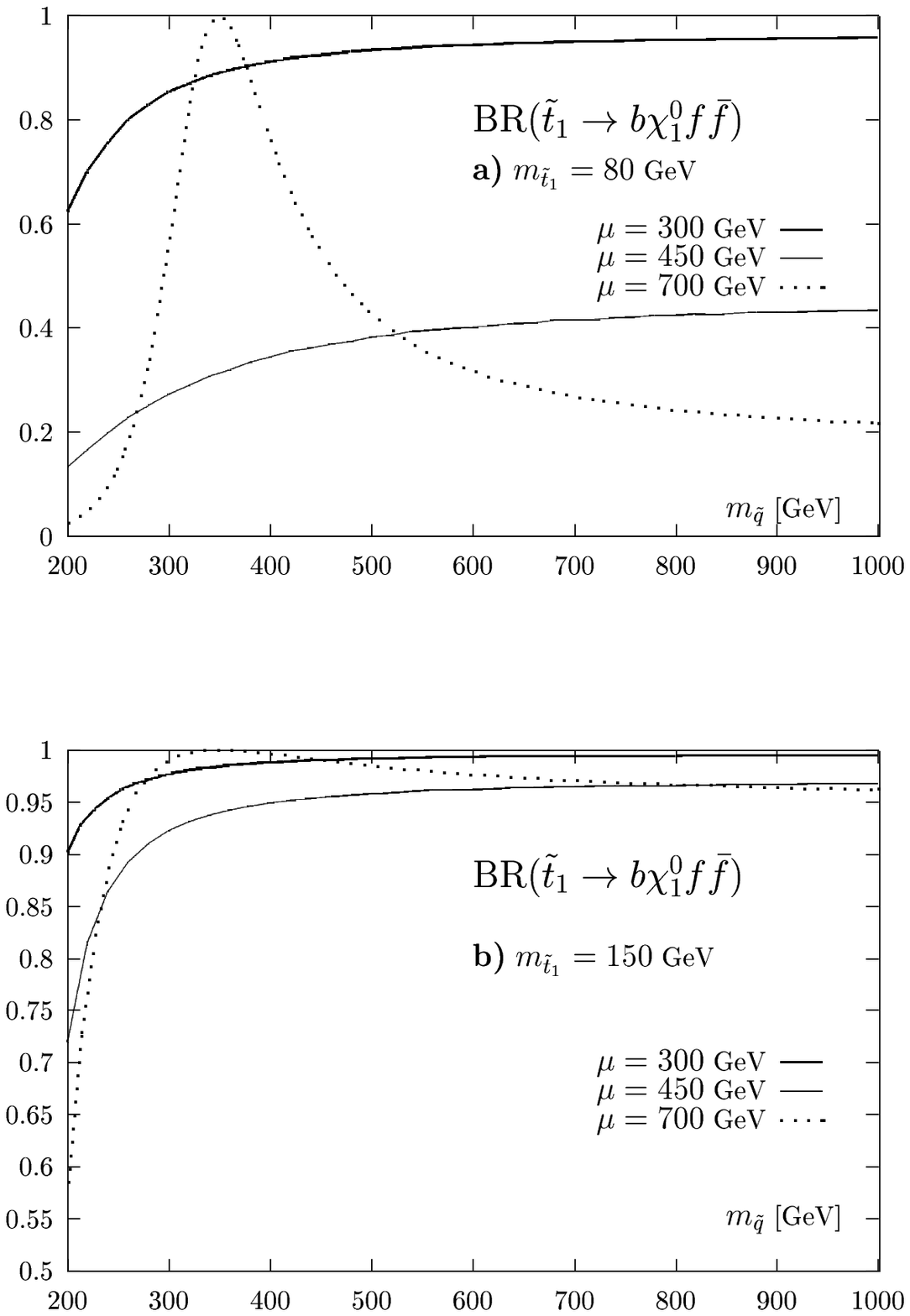,width=20cm}}
\vspace*{-7cm}
\caption[]{The branching ratio BR$(\tilde{t}_1 \ra b\chi_1^0 f \bar{f}')$ as
a function of the common squark mass $m_{\tilde{q}}$ for a scalar top mass
$m_{\tilde{t}_1}=80$ (a) and 150 GeV (b).}
\end{figure}
\newpage

\begin{figure}[htb]
\vspace*{-3.5cm}
\hspace*{-2cm}
\mbox{
\psfig{figure=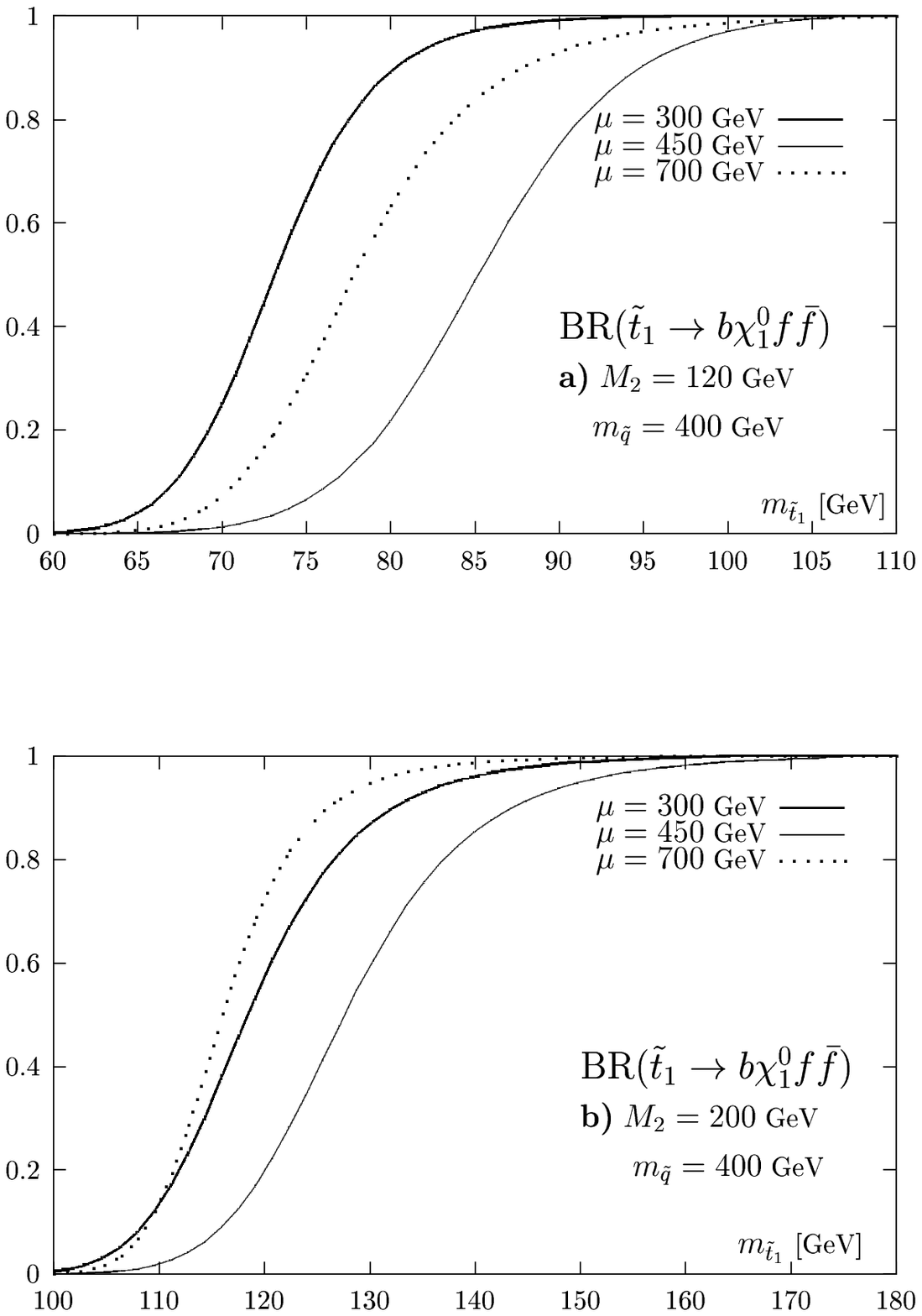,width=20cm}}
\vspace*{-7cm}
\caption[]{The branching ratio BR$(\tilde{t}_1 \ra b\chi_1^0 f \bar{f}')$ as
a function of the top squark mass $m_{\tilde{t}_1}$ for a scalar mass
$m_{\tilde{q}}=400$ GeV and a gaugino mass $M_2$ of 120 (a) and 200 GeV (b).}
\end{figure}
\newpage

\begin{figure}[htb]
\vspace*{-3.5cm}
\hspace*{-2cm}
\mbox{
\psfig{figure=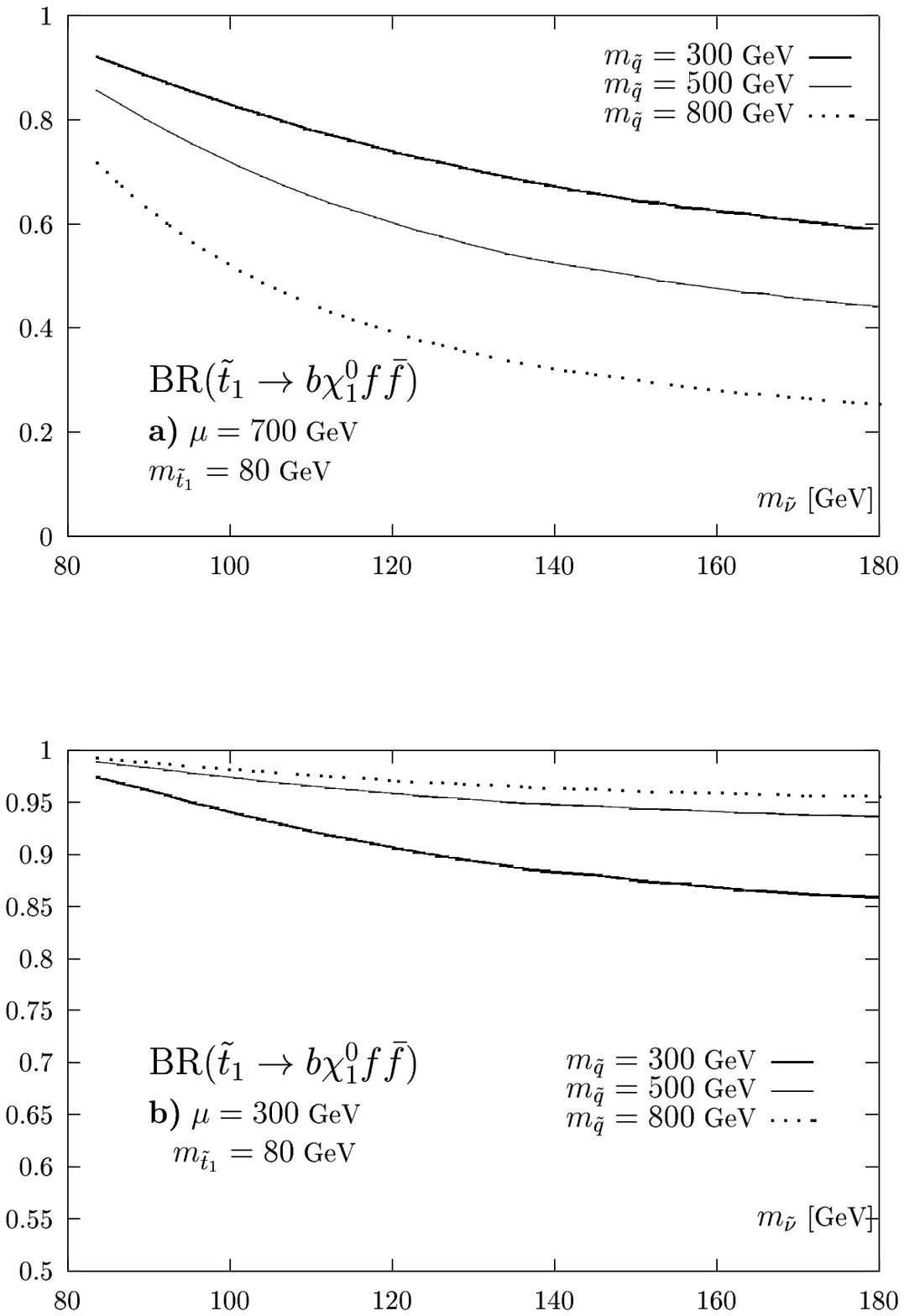,width=20cm}}
\vspace*{-7cm}
\caption[]{The branching ratio BR$(\tilde{t}_1 \ra b\chi_1^0 f \bar{f}')$ as
a function of the sneutrino mass $m_{\tilde{\nu}}$ for a scalar top mass
$m_{\tilde{t}_1}=80$ and two values of $\mu=700$ GeV (a) and 300 GeV (b).}
\end{figure}

\newpage

\begin{figure}[htb]
\vspace*{-3.5cm}
\hspace*{-2cm}
\mbox{
\psfig{figure=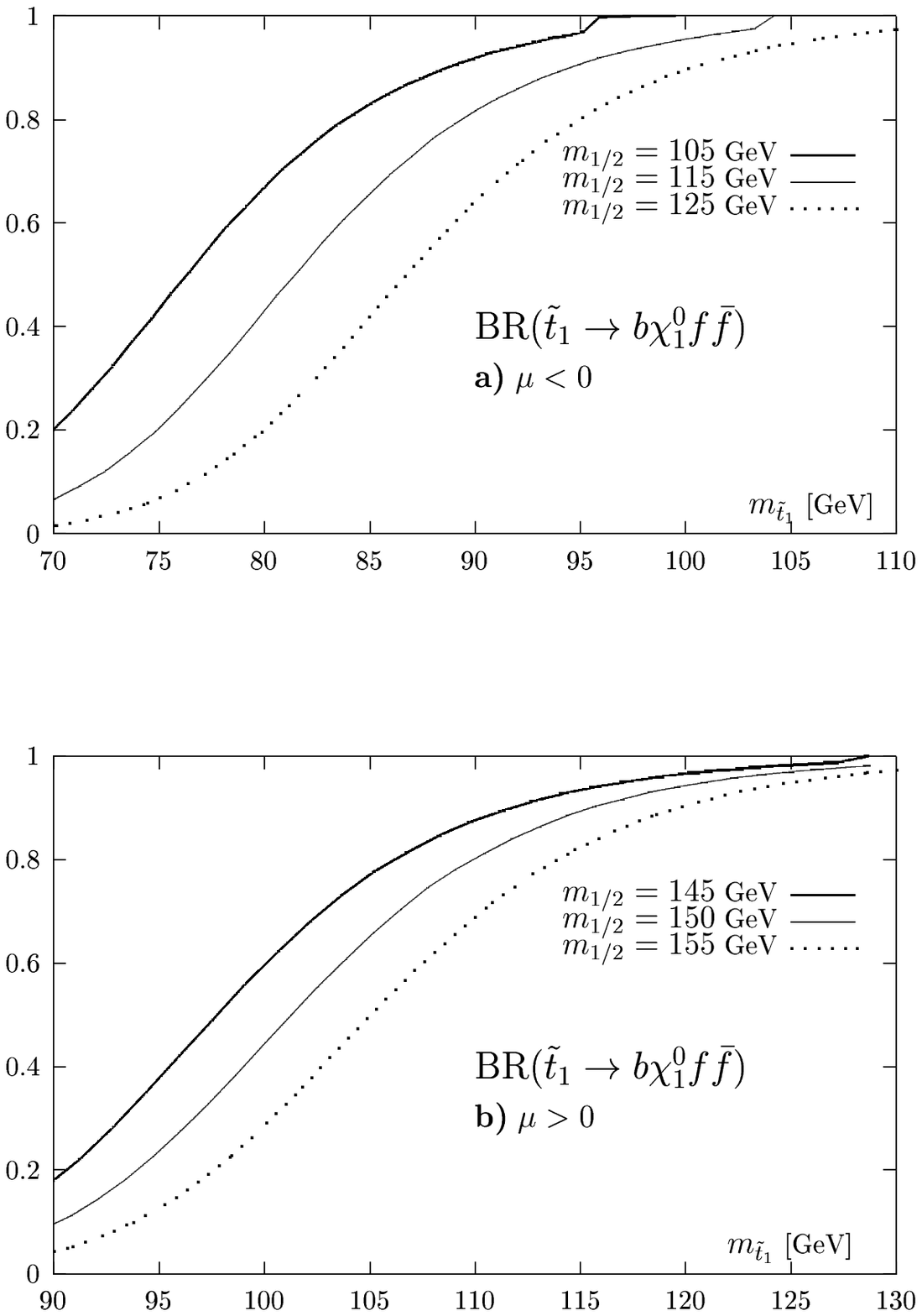,width=20cm}}
\vspace*{-7cm}
\caption[]{The branching ratio BR$(\tilde{t}_1 \ra b\chi_1^0 f \bar{f}')$ as
a function of stop mass $m_{\tilde{t}_1}$ in the mSUGRA scenario for 
$\mu<0$ (a) and $\mu>0$ (b).}
\end{figure}


\newpage

\begin{thebibliography}{99}

\bibitem{R1} For reviews on the MSSM, see: P. Fayet and S. Ferrara, Phys. 
Rep. 32 (1977) 249; H.P.~Nilles, Phys. Rep. 110 (1984) 1; R. Barbieri, 
Riv. Nuov. Cim. 11 (1988) 1; R. Arnowitt and Pran Nath, Report CTP-TAMU-52-93;
M. Drees and S. Martin, CLTP Report (1995) and hep-9504324; S. Martin, 
hep-ph9709356; J. Bagger, Lectures at TASI-95, hep-ph/9604232.  
%
\bibitem{R2} H. E. Haber and G. Kane, Phys. Rep. 117 (1985) 75. 
%
\bibitem{PDG} Particle Data Group, C.~Caso et al., Eur. Phys. Journal C3 
(1998) 1; for a more recent complilation see for instance, A. Djouadi et al., 
hep--ph/9901246. 
%
\bibitem{Tev} CDF Collaboration, Phys. Rev. D56 (1997) R1357; D0 Collaboration,
contribution to the EPS--HEP Conference, Jerusalem 1997, Ref.~102. 
%
\bibitem{Lep2} F. Cerutti, Report from the LEP SUSY working group, 
talk given on behalf of the LEP SUSY working group, LEPC, 15 sept. 1998.
%
\bibitem{qmix} J. Ellis and S. Rudaz, Phys. Lett. B128 (1983) 248;
M. Drees and K. Hikasa, Phys. Lett. B252 (1990) 127. 
%
\bibitem{twobod} For recent reviews of the two--body decays of top squarks, 
see A. Bartl et al., hep--ph/9709252; 
W. Porod,  hep-ph/9804208; S. Kraml, hep-ph/9903257; 
T. Plehn,  hep-ph/9809319. 
%
\bibitem{Rp} P. Fayet, Phys. Lett. 69B (1977) 489.
%
\bibitem{Hikasa} K.I. Hikasa and M. Kobayashi, Phys. Rev. D36 (1987) 724. 
%
\bibitem{ppstop} G. Kane and J.P. Leveille, Phys. Lett. 112B (1982) 227; 
P.R. Harrison and CH. Llewellyn--Smith, Nucl. Phys. B213 (1983) 223; 
E. Reya and DP. Roy, Phys. Rev. D32 (1985) 645; 
S. Dawson, E. Eichten and C. Quigg, Phys. Rev. D31 (1985) 1581; 
H. Baer and X. Tata, Phys. Lett. 160B (1985) 159;  
W. Beenakker, M. Kr\"amer, T. Plehn, M. Spira and 
P.M. Zerwas, Nucl. Phys. B515 (1998) 3. 
%
\bibitem{topdec} H. Baer and X. Tata, Phys. Lett. 167B (1986) 241; 
H. Baer, M. Drees, R. Godbole, J. Gunion and  X. Tata, Phys. Rev. D44 
(1991) 725;  M. Borzumati and N. Polonsky, hep-ph/9602433; 
A. Djouadi, W. Hollik  and C. J\"unger, Phys. Rev. D54 
(1996) 5629; C.S. Li, R. J. Oakes and J. M. Yang, Phys. Rev. D54 (1996) 6883.
%
\bibitem{CDFstop}
CDF Collaboration, Abstract 652, Conference ICHEP98, Vancouver, July 1998. 
%
\bibitem{eestop} 
For a review see, E. Accomando et al., Phys. Rept. 229 (1998) 1;
A. Bartl et al., Z. Phys. C76 (1997) 549; 
J. Schwinger, ``Particles, Sources and Fields", Addison-Wesley
Reading, MA, 1973; M. Drees and K. Hikasa, in Ref.~\cite{qmix}; 
A. Arhrib, M. Capdequi-Peyranere and A. Djouadi,
Phys. Rev. D52 (1995) 1404;  H. Eberl, A. Bartl and W. Majerotto,
Nucl. Phys. B472 (1996) 481; W. Beenakker, R. Hopker and P.M. 
Zerwas, Phys. Lett. B378 (1996) 159;
W. Beenakker, R. Hopker, T. Plehn and P.M. Zerwas,
Z. Phys. C75 (1997) 349. 
%
\bibitem{Opal} OPAL Collaboration, G. Abbiendi et al.,  Phys. Lett. B456 
(1999) 95. 
%
\bibitem{porod} W. Porod and T. W\"ohrmann, Phys. Rev. D55 (1997) 2907;
W. Porod, Phys. Rev. D59 (1999) 095009. 
%
\bibitem{4bodyold} 
G. Altarelli and R. R\"uckl, Phys. Lett. 144B (1984) 126;  
I. Bigi and S. Rudaz, Phys. Lett. 153B (1985) 335.
%
\bibitem{fortran} Y. Mambrini, Ph. D. Thesis, in preparation. 
%
\bibitem{GUTrelations} M. Carena et al.,   Nucl. Phys. B426 (1994) 269;
 W. de Boer, R. Ehret and D.I. Kazakov, Z. Phys. C67 (1994) 647; M. Drees 
and S. Martin in \cite{R1}. 
%
\bibitem{R10} J.F. Gunion and H.E. Haber, Nucl. Phys. B272 (1986) 1. 
%
\bibitem{leprho} See for instance, G. Altarelli, hep--ph/9611239;
J. Erler and P. Langacker, hep--ph/9809352; G.C. Cho et al., hep--ph/9901351.  
%
\bibitem{drho} 
M. Drees and K. Hagiwara, Phys. Rev. D42 (1990) 1709;
M.~Drees, K.~Hagiwara and A.~Yamada, Phys. Rev. D45  (1992) 1725; 
P.~Chankowski et al., Nucl. Phys.  B417 (1994) 101;
D.~Garcia and J.~Sol\`a, Mod.\ Phys.\ Lett.\ A9 (1994) 211;
A.~Djouadi et al., Phys. Rev. Lett. 78 (1997) 3626.
%

\bibitem{Rambo} R. Kleiss, W.J. Stirling and S.D. Ellis, Comput. Phys. 
Commun. 40 (1986) 359.

\end{thebibliography}
\end{document}